\documentclass[preprint,amssymb]{aastex631}

\usepackage{amsmath}
\renewcommand{\vec}[1]{\boldsymbol{#1}}

\def\cpc{{\itshape Comput. Phys. Commun.}}
\def\pop{{\itshape Phys. Plasmas}}

\begin{document}

\title{A simple procedure for generating a Kappa distribution in PIC simulation}

\author[0000-0002-0945-1815]{Seiji Zenitani}
\affiliation{Space Research Institute, Austrian Academy of Sciences, 8042 Graz, Austria}
\email{seiji.zenitani@oeaw.ac.at}

\begin{abstract}
For kinetic modeling of plasma processes in space,
a rejection-sampling procedure for generating a Kappa distribution in particle-in-cell (PIC) simulation is proposed. 
A Pareto distribution is employed as an envelope distribution.
The procedure only requires uniform variates,
and its acceptance efficiency is $\approx 0.73$--$0.8$.
\end{abstract}

\keywords{Plasma physics (2089) --- Computational methods (1965) --- Algorithms (1883) --- Monte Carlo methods (2238)}

\section{Introduction}

A Kappa distribution is one of the most important velocity distributions in space plasmas \citep{pierrard10,kappa}.
The distribution attracts growing attention,
because it is ubiquitously observed across the heliosphere, and
because it is related to a maximum state of non-Boltzmann entropy. 
Owing to its importance, there is a growing demand
for kinetic modeling of plasma processes with Kappa distributions \citep[e.g.,][]{park13,gedalin22,ma23}.
In kinetic simulations such as particle-in-cell (PIC) simulation,
it is usually necessary to initialize Kappa-distributed particles
by using random numbers. 

The Kappa distribution is equivalent to
a three-dimensional (trivariate) $t$-distribution.
A standard way for generating such a $t$-distribution is
to divide a multivariate normal distribution
by a gamma-distributed random number. 
Recent works \citep{abdul15,zeni22} employ this strategy
for generating a Kappa distribution in PIC simulation.
Unfortunately, gamma generators are not always provided in programming languages,
and so the user often needs to implement one of many gamma generators \citep{devroye86,luengo22}. 
This fact makes PIC codes with the standard Kappa generator less portable. 
There exist $t$-generators that only require uniform variates
\citep{bailey94,abdul14}, however,
they are only available for 1D and 2D distributions. 
In this research note, I propose
a random number generator for a 3D Kappa distribution
that only requires uniform variates.

\section{Kappa generator}

\begin{figure*}[htbp]
\centering
\includegraphics[width={\textwidth}]{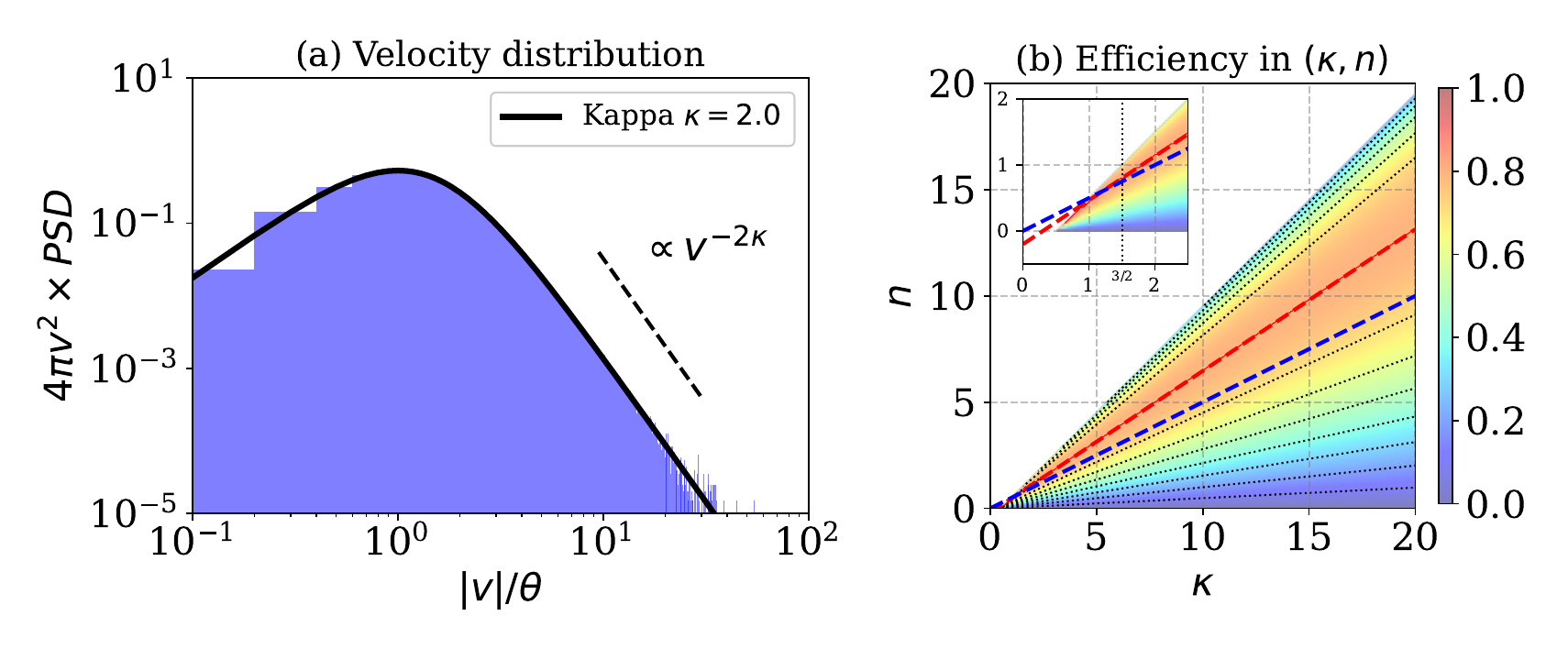}
\caption{
(a) Velocity distribution of the Kappa distribution with $\kappa=2$.
The black line and the blue histogram show
the theoretical curve and numerical results.
(b) Acceptance efficiency in the $(\kappa,n)$ space (Eq.~\eqref{eq:eff}).
The thin red line, the red dashed line, and the blue dashed line indicate
the optimum line,
the near-optimum approximation ($n= {2\kappa}/{3} - {1}/{5}$), and
my recommendation ($n = \kappa/2$), respectively.
\label{fig:kappa}}
\end{figure*}

The phase-space density of a Kappa distribution is given by
\begin{align}
\label{eq:kappa}
f_{\kappa}(\vec{v})d^3{v}
= \frac{N_0}{(\pi\kappa\theta^2)^{3/2}} \frac{\Gamma(\kappa+1)}{\Gamma(\kappa-1/2)}  \left( 1 + \frac{ \vec{v}^2 }{\kappa \theta^2} \right)^{-(\kappa+1)} d^3{v}
,
\end{align}
where $N_0$ is the density, $\kappa$ is the spectral index, $\theta$ is the characteristic speed,
and $\Gamma(x)$ is the gamma function.
Its omni-directional distribution with $\kappa=2$ is shown
by the black curve in Figure \ref{fig:kappa}(a).
It has a power-law tail, $\propto v^{-2\kappa}$.

Below I will construct a Monte Carlo procedure.
Considering a new variable
$x \equiv {v^2}/({\kappa\theta^2})$
and the spherical coordinates ($d^3v \rightarrow 4\pi v^2 dv$),
the distribution can be rewritten by
\begin{align}
\label{eq:kappa_x}
f(x)
\,dx
=
\frac{1}{\mathrm{B}\left( \frac{3}{2}, \kappa-\frac{1}{2} \right)}
x^{1/2}(1+x)^{-\kappa-1}
\,dx
,
\end{align}
where $\mathrm{B}\left( \alpha, \beta \right)$ is the beta function.
This distribution is a beta-prime distribution,
also known as a beta distribution of the second kind,
with shape $(3/2, \kappa-1/2)$.

I employ the Pareto distribution in $x \in (0, \infty)$
as an envelope function of a rejection sampling method:
\begin{align}
\label{eq:envelope}
g(x) \equiv n (1+x)^{-(n+1)}
.
\end{align}
The index $n$ ranges
$0 < n \le \kappa - \frac{1}{2}$
such that an area $\int_0^\infty g(x)dx$ can be defined, and that
$g(x) \propto x^{-(n+1)} $ decays slower than
$f(x) \propto x^{-\kappa-1/2}$ in the $x \rightarrow \infty$ limit.
The Pareto distribution (Eq.~\eqref{eq:envelope}) can be generated
by using a uniform variate $U_1 \sim U(0,1)$,
\begin{align}
x \leftarrow (1-U_1)^{-1/n} - 1
.
\label{eq:pareto}
\end{align}
If $U_1$ is defined in $(0,1]$,
one can replace $(1-U_1)$ with $U_1$
here and in subsequent discussions.

Then I consider the maximum value of $f(x)/g(x)$,
\begin{align}
c = \max\left( \frac{f(x)}{g(x)} \right)
= \frac{f(x_{\rm m})}{g(x_{\rm m})}
= \frac{\mathcal{D}}{n \, \mathrm{B}\left( \frac{3}{2},\kappa - \frac{1}{2} \right)},
\label{eq:rej_c}
\end{align}
where
\begin{align}
x_{\rm m} = \frac{1}{2\kappa-2n-1},
~~~
\label{eq:rej_c}
\mathcal{D}
&=
\left( {2\kappa-2n-1} \right)^{\kappa - n - 1/2}
\left( {2\kappa-2n} \right)^{n - \kappa}
.
\end{align}
Using them, I accept the output of Eq.~\eqref{eq:pareto},
when another uniform variate $U_2$ satisfies
\begin{align}
U_2
\le
\frac{f(x)}{c \cdot g(x)}
=
\frac{ x^{1/2}(1+x)^{n-\kappa} }{\mathcal{D}}
.
\label{eq:rej}
\end{align}
Then I obtain $x$ that follows the beta-prime distribution (Eq.~\eqref{eq:kappa_x}).
The acceptance efficiency is
\begin{align}
\mathrm{Eff}(\kappa,n)
=
c^{-1}
=
\frac{n}{\mathcal{D}}
\mathrm{B}\left( \frac{3}{2},\kappa - \frac{1}{2} \right)
\label{eq:eff}
.
\end{align}

After sampling $x$, I translate it to $|v|$, and then
randomly scatter the vector into 3D directions.
A full procedure is shown below.
Step 0 is necessary only once, while other steps are required for each particle.
Fortunately, the beta function does not appear in the procedure. 
\begin{itemize}
\item
{\bf Step 0 (initialization):}
$\mathcal{D} \leftarrow
\left( {2\kappa-2n-1} \right)^{\kappa-n-1/2}
\left( {2\kappa-2n} \right)^{n-\kappa}$
\item
{\bf Step 1:}
Generate $U_1, U_2 \sim U(0, 1)$
\item
{\bf Step 2:}
$W \leftarrow \sqrt{(1-U_1)^{-{1}/{n}}-1}$
\item
{\bf Step 3:}
If $W (1-U_1)^{\frac{\kappa-n}{n}} < \mathcal{D} U_2 $, go to Step 1.
\item
{\bf Step 4:}
Generate $U_3, U_4 \sim U(0, 1)$ 
\item
{\bf Step 5:}
$V \leftarrow \sqrt{ \kappa\theta^2 } W$,~
$v_x \leftarrow V ( 2 U_3 - 1 )$,~
$v_y \leftarrow 2 V \sqrt{ U_3 (1-U_3) } \cos(2\pi U_4)$,~
$v_z \leftarrow 2 V \sqrt{ U_3 (1-U_3) } \sin(2\pi U_4)$
\end{itemize}

The index $n$ needs to be set separately. 
I have carried out a grid search of the acceptance efficiency (Eq.~\eqref{eq:eff}) in the $(\kappa,n)$ space in Figure \ref{fig:kappa}(b).
The optimum value is indicated by the thin red line. 
The inlet focuses on a key region near the origin.
Note that the standard Kappa distribution is defined for $\kappa > 3/2$,
to the right of the vertical dotted line in the inlet.

I approximate $n = a\kappa + b$
to estimate the efficiency (Eq.~\eqref{eq:eff})
in the $\kappa \rightarrow \infty$ limit. 
For $a = 1$,
with help from
the relation of the beta function for a large $\beta$,
i.e., $\mathrm{B}(\alpha,\beta) \approx \Gamma(\alpha) \beta^{-\alpha}$,
I find that the efficiency decays to zero,
\begin{align}
\mathrm{Eff}(\kappa,n)
&\approx
{\kappa \Gamma\left(\frac{3}{2}\right) \kappa^{-3/2} }
\Big( -2b-1 \Big)^{b+1/2} \nonumber
\left( {-2b} \right)^{ -b } 
~
\propto \kappa^{-1/2}
\rightarrow 0
.
\end{align}
For $a\ne 1$,
with help from $\lim_{k \rightarrow \infty} \left( 1+ \frac{x}{k} \right)^k = \exp(x)$,
the asymptotic efficiency is given by
\begin{align}
\lim_{\kappa \rightarrow \infty}
\mathrm{Eff}(\kappa,n)
& \approx
{a\kappa \Gamma\left(\frac{3}{2}\right) \kappa^{-3/2} }
\Big( 2(1-a)\kappa \Big)^{1/2}
\times
\lim_{\kappa \rightarrow \infty}
\left( 1+ \frac{1}{2(1-a)\kappa} \right)^{ (1-a)\kappa }
\approx a \frac{\sqrt{2\pi e (1-a)}}{2}
\label{eq:asymptotic}
.
\end{align}
The asymptotic value only depends on $a$,
in agreement with the radially similar profile in Figure \ref{fig:kappa}(b). 
The zero-coefficient $b$ is important to adjust the efficiency near the origin,
as seen in the inlet in Figure \ref{fig:kappa}(b).

I propose two approximations for $n$.
The first one, $n= {2\kappa}/{3} - {1}/{5}$,
is indicated by the red dashed line in Figure \ref{fig:kappa}(b). 
This is obtained by fitting the optimum line. 
This near-optimum approximation gives $\approx 0.8$ regardless of $\kappa$. 
In fact, $a=2/3$ maximizes the asymptotic efficiency,
${\sqrt{2 \pi e}}/({3\sqrt{3}}) \approx 0.795$ (Eq.~\eqref{eq:asymptotic}). 

The second one, $n = \kappa/2$,
is indicated by the blue dashed line in Figure \ref{fig:kappa}(b). 
This is my recommendation.
Although it is slightly less efficient than the first one,
it gives $\approx 0.73$--$0.8$.
It starts from $0.806$ at $\kappa=1.5$, $0.785$ at $\kappa=2$, $0.750$ at $\kappa=5$, and then it is asymptotic to ${ \sqrt{\pi e} }/{ 4 } \approx 0.731$ (Eq.~\eqref{eq:asymptotic}).
Interestingly, $n = \kappa/2$ simplifies exponents in Steps 2 and 3 to $-2/\kappa$ and unity, which makes the code faster.
Step 0 is also $\mathcal{D} \leftarrow \sqrt{ (\kappa-1)^{\kappa-1} / \kappa^\kappa }$.

With $n = \kappa/2$,
I have generated a Kappa distribution with $\kappa=2$
by using $10^6$ particles.
The numerical results are shown in the blue histogram
in Figure \ref{fig:kappa}(a),
in agreement with the theory (the black curve).

\section{Discussion}

I have proposed a rejection-sampling procedure
for generating a Kappa distribution in PIC simulation. 
It only uses uniform variates, and it should be highly portable.

The total number of random variates is
one of the key factors in controlling computational cost. 
The standard Kappa generator \citep{abdul15,zeni22}
requires three normal variates and one gamma variate per particle.
Since a popular gamma generator by \citet{mt00} calls
one normal variate and one uniform variate, and
since its efficiency is $\approx 1$,
the standard Kappa generator requires
four normal variates and one uniform variate per particle. 
Considering its efficiency of $\approx 0.73$--$0.8$,
the proposed method calls $\approx 4.5$--$4.7$ uniform variates per particle. 
Therefore, it would be computationally less expensive than the standard method.

\end{document}